\documentclass[doublecol]{epl2} 

\title{Time-separated oscillatory fields for high-precision mass measurements on short-lived Al and Ca nuclides}
\shorttitle{High-precision mass measurements} 

\author{S. George\inst{1,2}\thanks{E-mail: \email{george@uni-mainz.de}}
\and G. Audi\inst{3} 
\and B. Blank\inst{4}
\and K. Blaum\inst{1,2,5} 
\and M. Breitenfeldt\inst{6}
\and U. Hager\inst{7}\thanks{present address: TRIUMF, Vancouver, B.C., V6T, 2A3}
\and F. Herfurth\inst{1}
\and A. Herlert\inst{8} 
\and A. Kellerbauer\inst{5}
\and H.-J. Kluge\inst{1,9} 
\and M. Kretzschmar\inst{2} 
\and D. Lunney\inst{3}
\and R. Savreux\inst{1}
\and S. Schwarz\inst{10}
\and L. Schweikhard\inst{6} 
\and C. Yazidjian\inst{1}
}

\shortauthor{S. George \etal}

\institute{                    
  \inst{1} GSI, Planckstra{\ss}e 1, 64291 Darmstadt, Germany\\
  \inst{2} Institut f\"ur Physik, Johannes Gutenberg-Universit\"at, 55099 Mainz, Germany\\
  \inst{3} CSNSM-IN2P3-CNRS, 91405 Orsay-Campus, France\\
  \inst{4} Centre d'Etudes Nucl$\acute{\mbox{e}}$aires de Bordeaux-Gradignan, 33175
           Gradignan Cedex, France\\
  \inst{5} Max-Planck-Institut f\"ur Kernphysik, 69117 Heidelberg, Germany\\
  \inst{6} Institut f\"ur Physik, Ernst-Moritz-Arndt-Universit\"at, 17487 Greifswald, Germany\\
  \inst{7} Department of Physics, University of Jyv\"askyl\"a, P.O. Box 35 (YFL), 40014
           Jyv\"askyl\"a, Finland\\
  \inst{8} CERN, Physics Department, 1211 Geneva 23, Switzerland\\
  \inst{9} Physikalisches Institut, Ruprecht-Karls-Universit\"at, 69120 Heidelberg, Germany\\
  \inst{10} NSCL, Michigan State University, East Lansing, MI 48824-1321, USA\     
}


\abstract{
High-precision Penning trap mass measurements on the stable nuclide $^{27}$Al as well as on the short-lived radionuclides $^{26}$Al and $^{38,39}$Ca have been performed by use of radiofrequency excitation with time-separated oscillatory fields, i.e. Ramsey's method, as recently introduced for the excitation of the ion motion in a Penning trap, was applied. A comparison with the conventional method of a single continuous excitation demonstrates its advantage of up to ten times shorter measurements. The new mass values of $^{26,27}$Al clarify conflicting data in this specific mass region. In addition, the resulting mass values of the superallowed $\beta$-emitter $^{38}$Ca as well as of the groundstate of the $\beta$-emitter $^{26}$Al$^\mathit{m}$ confirm previous measurements and corresponding theoretical corrections of the \textit{f}\textit{t}-values.}

\begin{document}

\maketitle

\section{Introduction}
Superallowed $0^+\rightarrow 0^+$ $\beta$-decays are sensitive probes for testing fundamental concepts of weak interaction. Hardy and Towner~\cite{Hard2005a} addressed 20 such decays  covering the nuclear chart from $^{10}$C to $^{74}$Rb and showed a consistent picture allowing a confirmation of the conserved-vector-current (CVC) hypothesis. Three properties of the nuclear transitions are merged in the comparative half-life \textit{f}\textit{t}: the transition energy $Q_{EC}$, the half-life of the mother nuclei $t_{1/2}$, and the branching ratio $R$. Since the Fermi-type $\beta$-decays are only effected by the vector part of the hadronic weak interaction, the CVC hypothesis claims identical \textit{f}\textit{t}-values for all transitions between isospin $T=1$ analog states irrespective of theoretical corrections. Including these modifications the corrected \textit{f}\textit{t}-value can be written in the form
\begin{eqnarray}
\label{eq.1}
\mathcal{F}t & \equiv & ft(1+\delta_R')(1+\delta_{NS}-\delta_C)\nonumber\\
& = & \frac{K}{2G_V^2(1+\Delta_R^V)}=const.,
\end{eqnarray}
where $K$ is a numerical constant and $G_V$ is the vector coupling constant. $\delta_C$ denotes the isospin-symmetry-breaking correction, $\delta_R'$ and $\delta_{NS}$ are the transition-dependent parts of the radiative correction, while $\Delta_R^V$ is transition-independent~\cite{Hard2002}. Moreover, $\delta_C$ and $\delta_{NS}$ depend on the nuclear structure of the specific nuclei. In~\cite{Hard2005a} the averaged $\mathcal{F}$\textit{t}-value of the 12 best known-cases was determined to be
\begin{equation}
\label{eq.2}
\overline{\mathcal{F}t}=3072.7\pm 0.8\mbox{~s} .
\end{equation}
\\
\begin{figure}[h!]
    \centering
       \includegraphics[width=8cm]{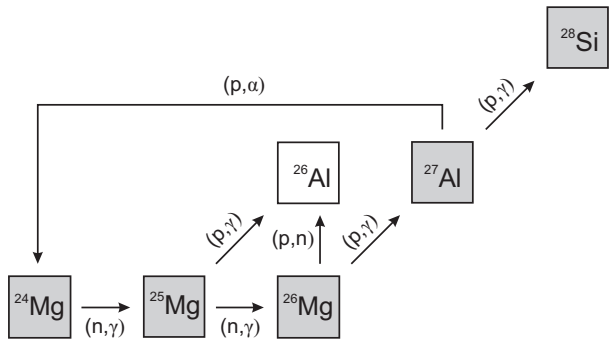}
        \caption{Section of the nuclear chart in the region of interest for the Al mass measurements indicating the reactions connecting the nuclides. Grey boxes denote stable isotopes.} 
        \label{AL26-Massenregion}
\end{figure}

Recently Savard \textit{et al.}~\cite{Sava2005} published a new $Q_{EC}$ value of the superallowed decay of $^{46}$V measured with the Canadian Penning trap facility. Their $Q$-value differed by 2.19 keV corresponding to 2.5 standard deviations from a combination of two contradictory data~\cite{Hard2005a,Squi1976,Vona1977}. The high $\mathcal{F}t$-value that resulted was conspicious with respect to the average $\overline{\mathcal{F}t}$-value of the 12 best-known superallowed transitions. Together with an updated data set of these 12 transition it is leading to a new average value of $\overline{\mathcal{F}t}=3073.66(75)$\,s. This triggered a careful remeasurement of the Canadian Penning trap result as well as a search for possible systematic differences between Penning trap measurements and $Q$-values from reaction measurements as performed by Hardy and coworkers~\cite{Hard2006}. The result of Savard \textit{et al.} has meanwhile been confirmed by the Penning trap facility JYFLTRAP~\cite{Eron2006}. Obviously, there is a strong demand to remeasure the reaction $Q$-values of superallowed decays with high-precision Penning trap facilities.

The Penning trap mass spectrometer ISOLTRAP has previously already addressed three superallowed emitters of the 12 best-known cases: $^{22}$Mg~\cite{Mukh2004}, $^{34}$Ar~\cite{Herf2002}, and $^{74}$Rb~\cite{Kell2004}. In the present paper, we contribute with mass measurements of $^{26}$Al, itself a daughter in a superallowed decay, and $^{38}$Ca. For the mass measurements Ramsey's method of separated oscillatory fields for excitation of the ion motion in a Penning trap was applied. In ~\cite{Kret2007,Geor2007a,Geor2007b,Suho2007,Eron2007} the adaption of Ramsey's method to Penning trap mass spectrometry has been derived and experimentally demonstrated. The precision of the frequency determination could be improved by more than a factor of three compared to the conventional excitation method and scan procedure. This allows one to reach a given precision up to ten times faster than before. 
\\
\begin{figure}[h!]
    \centering
       \includegraphics[width=8cm]{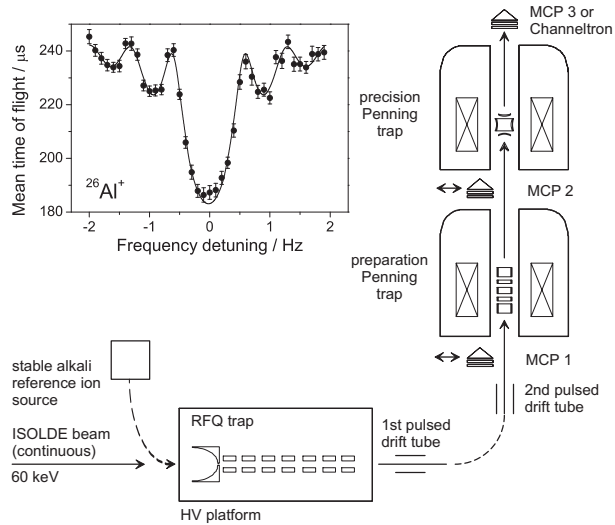}
        \caption{Experimental setup of the mass spectrometer ISOLTRAP. The RFQ trap, the preparation and precision Penning traps as well as the reference ion source are shown. Monitoring the ion transfer as well as the time-of-flight measurements for the determination of the cyclotron frequency are done with micro-channel-plate (MCP) detectors or a channeltron~\cite{Yazi2007}. In the inset a cyclotron frequency resonance for $^{26}$Al ions with an excitation time of 1.5\,s is given.} 
        \label{ISOLTRAP_Setup}
\end{figure}
Furthermore, the direct mass measurements of $^{26}$Al and $^{27}$Al address a problem of conflicting data derived from various reaction $Q$-value measurements~\cite{Waps2003}: The masses of the stable nuclides $^{24,26}$Mg~\cite{Berg2003} and $^{28}$Si~\cite{Difi1994,Jert1993} have all been measured with Penning trap facilities. They are also related via reaction $Q$-values, namely $^{24,26}$Mg by a pair of (n,$\gamma$) reactions through the stable isotopes $^{25}$Mg and $^{26}$Mg, and $^{28}$Si by a pair of (p,$\gamma$) reactions through the isotope $^{27}$Al. In addition, the masses of $^{25}$Mg and $^{26}$Mg are also related by a sequence of a  (p,$\gamma$) reaction and a (p,n) reaction through the isotope $^{26}$Al (see fig.~\ref{AL26-Massenregion}). The two most precise values of the $^{25}$Mg(n,$\gamma$)$^{26}$Mg reaction agree neither with each other nor with the results from the combined $^{25}$Mg(p,$\gamma$)$^{26}$Al and $^{26}$Mg(p,n)$^{26}$Al reactions. A similar inconsistency occurs for the nuclide $^{27}$Al in relation to the magnesium isotopes and $^{28}$Si. Thus we performed direct mass measurements of $^{26,27}$Al to resolve this unsatisfactory situation.
\section{Experimental Setup and measurement procedure}
All measurements reported here have been performed with the triple-trap mass spectrometer ISOLTRAP~\cite{Herf2003,Blau2005,Mukh2007} (see fig.~\ref{ISOLTRAP_Setup}) at the online mass separator ISOLDE/CERN~\cite{Kugl2000}. The experimental procedure is as follows: The 60-keV continuous ISOLDE ion beam is accumulated in a linear gas-filled RFQ ion beam cooler and buncher~\cite{Herf2001}. After a few milliseconds the accumulated ions are transferred in a bunch to the first Penning trap~\cite{Boll1996}, which is used for mass-selective buffer-gas cooling~\cite{Sava1991} to remove isobaric contaminations. In the second Penning trap the actual mass measurements are performed. $^{39}$K$^+$ ions from the stable alkali reference ion source for the measurements of the calcium isotopes and $^{23}$Na$^+$ ions from ISOLDE for the aluminium isotopes were used for calibration of the magnetic field strength.
\\
\begin{figure}[h!]
    \centering
        \includegraphics[width=8.5cm]{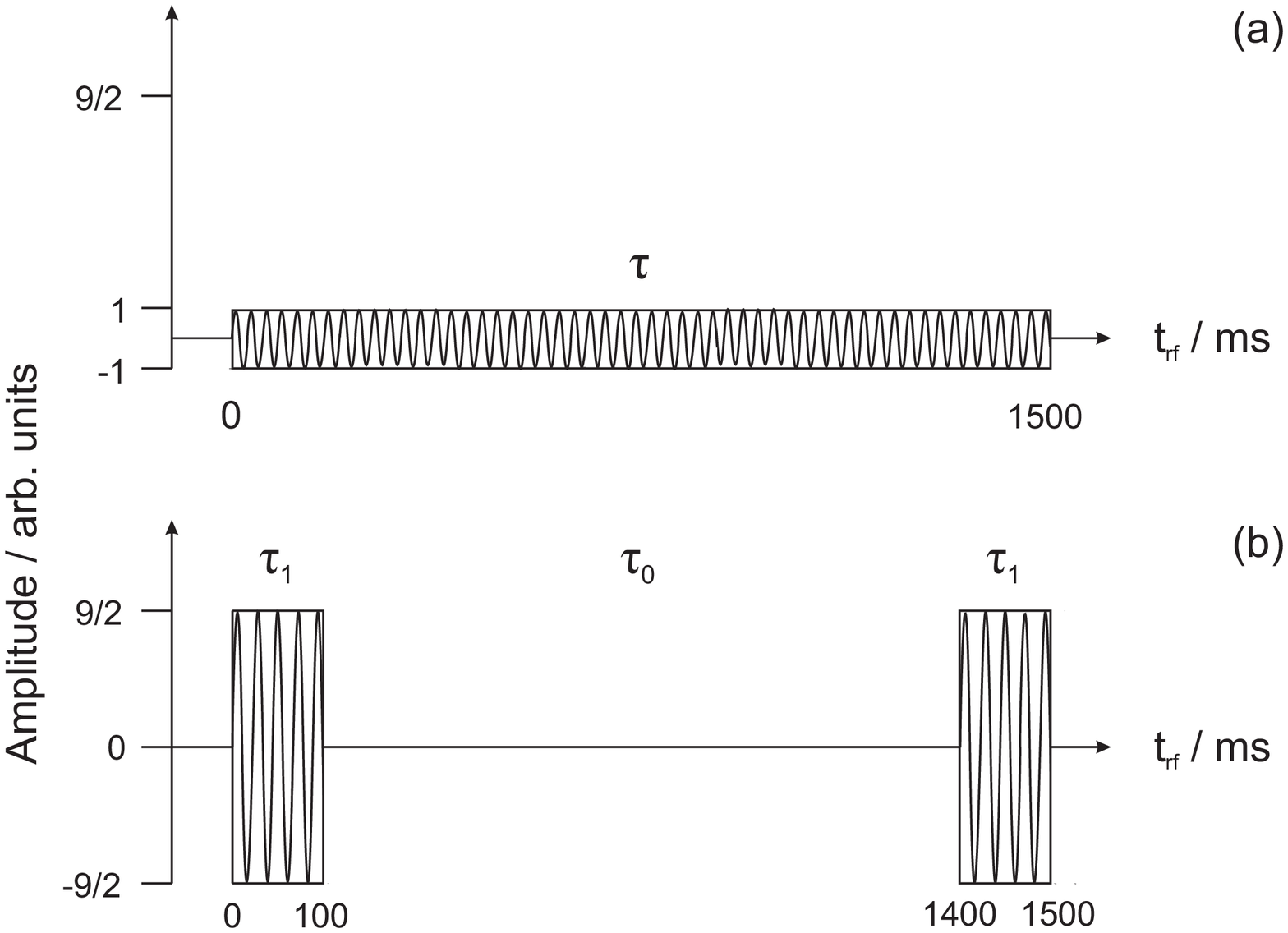}
        \caption{(a) Conventional excitation scheme with a continuous rf pulse of $\tau$=1.5\,s, (b) excitation with two 100-ms Ramsey pulses $\tau_1$ interrupted by a $\tau_0$=1.3\,s waiting period.} 
        \label{2Pulse}
\end{figure}
The measurements are based on the determination of the ions cyclotron frequency $\nu_c=qB/(2\pi m)$, which is proportional to the charge-to-mass ratio $q/m$ as well as to the magnetic field strength $B$~\cite{Blau2006}. The ion motion is excited by an azimuthal quadrupolar radiofrequency field via the four-fold segmented ring electrode of the Penning trap~\cite{Boll1992} before the ions are extracted and transported to a detector. The frequency of the excitation field is varied around the expected value to record a time-of-flight ion cyclotron resonance (TOF-ICR)~\cite{Gräf1980} (see inset of fig.~\ref{ISOLTRAP_Setup}) a well-established method in use at several facilities~\cite{Schw2006}.

The aluminium isotopes were produced by bombarding a silicon carbide target with $3\cdot10^{13}$ 1.4-GeV protons per pulse from the CERN proton synchrotron booster. A hot plasma source ionized the atoms released from the heated target. Isotopic separation was performed with the general purpose separator~\cite{Kugl2000} with a resolving power of typically 1000. 

For the calcium isotopes a heated titanium foil target was used in combination with a hot tungsten surface for ionization. The high-resolution separator~\cite{Kugl2000} served for mass separation with a resolving power of about 3000. In order to suppress isobaric contaminations by $^{38}$K ions, CF$_4$ was added in the ISOLDE ion source and the ions were delivered to ISOLTRAP in form of the molecular sidebands $^{38}$Ca$^{19}$F and $^{39}$Ca$^{19}$F. The mass of the ion is obtained by measuring the frequency ratio to a well-known reference ion.
\begin{figure}[h!]
    \centering
        \includegraphics[width=8cm]{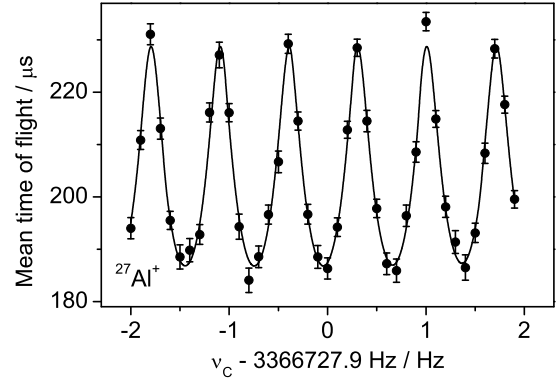}
        \caption{Time-of-flight cyclotron resonance of $^{27}$Al$^+$. A Ramsey excitation scheme was chosen with two excitation pulses of 100\,ms duration interrupted by a 1.3\,s waiting period. The solid line is a fit of the theoretically expected line shape to the data~\cite{Kret2007,Geor2007b}.} \label{RamseyResonance}
\end{figure}

\section{The Ramsey technique}

In order to improve the precision of the frequency determination, separated oscillatory fields, as introduced by Ramsey to nuclear-magnetic-resonance experiments, were adopted ~\cite{Kret2007,Geor2007a} and used for the first time in an online mass measurement~\cite{Geor2007b}. Instead of exposing the ions continuously to the external radiofrequency field~\cite{Blau2003}, two excitation pulses interrupted by a waiting period are used to excite the radial ion motion (see fig.~\ref{2Pulse}). The shape of the TOF-ICR curve (see fig.~\ref{RamseyResonance}) differs significantly from the conventional one (see inset of fig.~\ref{ISOLTRAP_Setup}). The smaller width of the central resonance peak and the more pronounced sidebands allow a more precise or faster frequency measurements~\cite{Geor2007a,Geor2007b}. 

\begin{table}
\caption{Ratios of the cyclotron frequencies of the isotopes investigated in this work.}
\label{TableRatios}
\begin{center}
\begin{tabular}{lrr}
\hline\hline& & \\*[-9pt]
Ion  & Reference & Frequency ratio,$\frac{\nu_{ion}}{\nu_{ref}}$\\& & \\*[-9pt]
\hline
& & \\*[-8pt]
$^{26}$Al & $^{23}$Na & 1.1303707761(104)
\\
$^{27}$Al & $^{23}$Na & 1.1736365541(108)
\\
$^{38}$Ca$^{19}$F & $^{39}$K & 1.4622576087(162)
\\
$^{39}$Ca$^{19}$F & $^{39}$K & 1.4877789303(165)
\\
\hline\hline
\end{tabular}
\end{center}
\end{table}

\begin{figure}[h!]
    \centering
        \includegraphics[width=8.5cm]{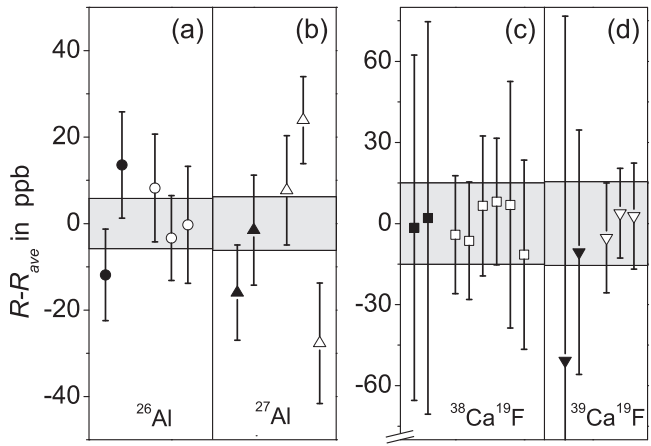}
        \caption{Difference between the measured cyclotron frequency ratios $R$ and their average value $R_{ave}$. Each data point contains the same number of recorded ions. Data points with filled symbols have been taken with the conventional excitation method, while empty symbols denote data points where the Ramsey method was used. In (a,b) the differences for the ratios $R$-$R_{ave}$ for $^{26,27}$Al$^+$ to $^{23}$Na$^+$ are shown. In (c,d) the same is plotted for the ratios of the molecular sidebands $^{38,39}$Ca$^{19}$F$^+$ compared to $^{39}$K$^+$. The grey band indicates the uncertainty of the average ratio.} 
        \label{Ratio}
\end{figure}

Penning trap mass measurements on the isotopes $^{26,27}$Al as well as on 
$^{38}$Ca$^{19}$F$^+$ and $^{39}$Ca$^{19}$F$^+$ have been performed with conventional excitation of the ion motion as well as with Ramsey's method of separated oscillatory fields. The number of recorded ion events per resonance is approximately 3000, which allows a comparison of systematic uncertainties between the two methods as well as a demonstration of the advantage of the new technique. In fig.~\ref{Ratio} the cyclotron-frequency ratios and their uncertainties are shown (see also tab.~\ref{TableRatios}). For simplicity the average ratio of each ion species is subtracted. Measurements with the conventional excitation (filled symbols) and with Ramsey-type excitation (empty symbols) can be directly compared. The total excitation time used for the aluminium isotopes was 1.5\,s. During the Ramsey-type excitation two 100 ms excitation pulses were interrupted by a 1.3\,s waiting period, so that the total excitation cycle remained 1.5\,s. 

\begin{table}
\caption{Mass excesses of the measured nuclides.}
\label{TableMassExcess}
\begin{center}
\begin{tabular}{lcr}
\hline\hline& \\*[-9pt]
Ion  & $\mbox{T}_{1/2}$ & Mass excess / keV \\
\hline
& \\*[-8pt]
$^{26}$Al & 717\,(24)\,ky & -12210.18 (22)
\\
$^{27}$Al & \mbox{stable} & -17196.92 (23)
\\
$^{38}$Ca & 440\,(8)\,ms & -22058.11 (60)
\\
$^{39}$Ca & 859.6\,(1.4)\,ms & -27282.59 (61)
\\\hline\hline
\end{tabular}
\end{center}
\end{table}
Due to the short half-life of $^{39}$Ca ($T_{1/2}$ = 859.6(1.4)\,ms) the excitation time was reduced to 1.2\,s and for $^{38}$Ca ($T_{1/2}$ = 440(8)\,ms) to 900\,ms, respectively. As before the duration of the Ramsey excitation pulses was 100\,ms and the total cycle length has been kept constant. 

\begin{figure}[h!]
    \centering
        \includegraphics[width=8cm]{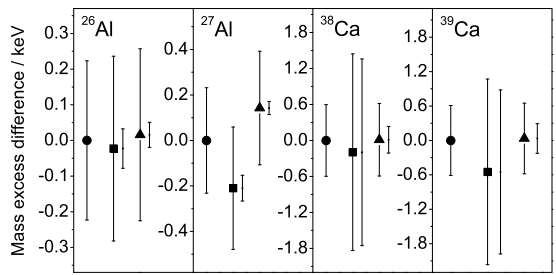}
        \caption{Mass excesses of all four radionucides derived from the ISOLTRAP measurements (full dots). In addition the mass excesses derived from measurements using the conventional excitation method (squares) as well as using the Ramsey excitation (triangles) are presented. The error bars next to the data points represent the statistical uncertainties.}
        \label{MassExcess}
\end{figure}

First, for all ion species, no significant difference of the extracted frequency ratios (fig.~\ref{Ratio}) and thus for the derived mass excess values (fig.~\ref{MassExcess}) is found between conventional and Ramsey excitation. All final mass excess values with their total uncertainties derived from the ISOLTRAP measurements are presented in tab.~\ref{TableMassExcess}. Second, the statistical uncertainties of the $^{38,39}$Ca measurements, which are less than $2\cdot 10^{-9}$ (shown separately in fig.~\ref{MassExcess}), show clearly that with Ramsey-type excitation only the systematic uncertainties are limiting the precision of the mass values. The systematic uncertainties are dominated by undetected magnetic field drifts during the measurements~\cite{Kell2003}. However, there is no difference in the statistical uncertainty between the continuous 1.5\,s excitation and a Ramsey-type excitation of equal length in the measurements of the aluminium isotopes. Thus, the Ramsey-type excitation is favorable for shorter excitation cycles: It allows for a 900\,ms excitation a frequency determination which is three times more precise than using the conventional excitation. This gain in precision in dependence of the overall excitation cycle has been observed for a constant number of collected ions~\cite{Geor2007a}, comparable to the data of the mass measurements presented here.
\section{The masses of $^{26}$Al and $^{27}$Al}
\begin{figure}[h!]
    \centering
        \includegraphics[width=8cm]{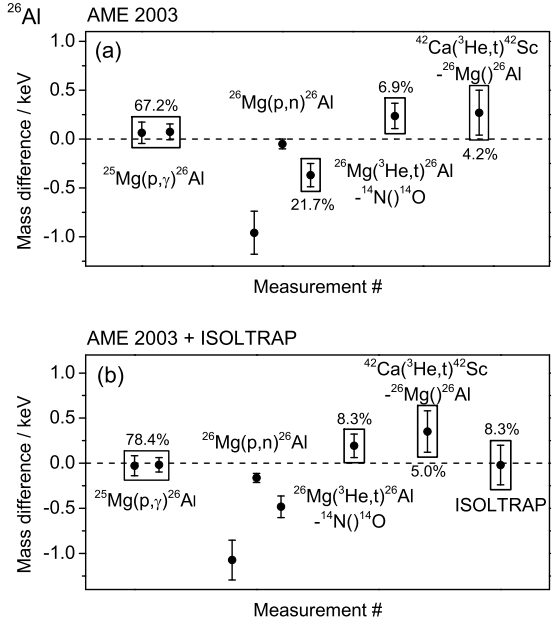}
        \caption{(a) Mass difference of $^{26}$Al between various reaction experiments and the average value of the atomic-mass evaluation from 2003~\cite{Audi2003}, here denoted as zero line. The boxes mark the measurements which contribute to the accepted value. The impact of the measurements is reflected by the percentage of the contribution to the final value. (b) Same as graph (a) but including the recent ISOLTRAP measurement. Note that the zero line is shifted due to the new average value in the new atomic-mass evaluation.}
         \label{Aluminium_26}
\end{figure}

Five different measurements of four types of reactions are contributing to the mass value of $^{26}$Al as published in the Atomic-Mass Evaluation AME2003~\cite{Audi2003}. All of them are shown in fig.~\ref{Aluminium_26}\,(a). The highest impact on the final AME2003 mass have two independent measurements of the reaction $^{25}$Mg(p,$\gamma$)$^{26}$Al~\cite{Berg1985,Kiks1991}. In addition $Q$-values of the reactions $^{26}$Mg(p,n)$^{26}$Al~\cite{Brin1994}, $^{26}$Mg($^3$He,t)$^{26}$Al-$^{14}$N()$^{14}$O~\cite{Kosl1987}, and $^{42}$Ca($^3$He,t)$^{42}$Sc-$^{26}$Mg()$^{26}$Al~\cite{Kosl1987} contribute to the mass of $^{26}$Al. The value derived from the reaction $^{26}$Mg(p,n)$^{26}$Al disagrees with former measurements of the same group~\cite{Bark1984,Bark1992}, as well as also disagrees with the four other measurements by more than three standard deviations. 

\begin{figure}[h!]
    \centering
        \includegraphics[width=8cm]{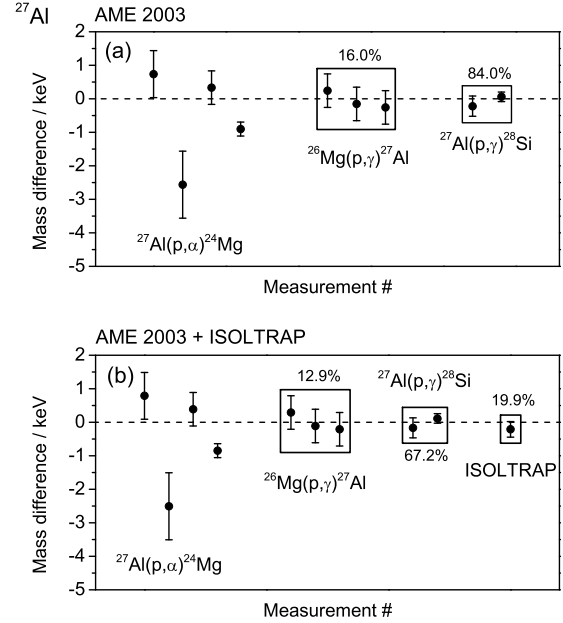}
        \caption{Same as fig.~\ref{Aluminium_26} but for $^{27}$Al.}
         \label{Aluminium_27}
\end{figure}

A new atomic-mass evaluation was performed with the ISOLTRAP data determined in this work which are the first derived from a Penning trap mass measurement. The data from the $^{26}$Mg(p,n)$^{26}$Al reaction is not taken into account anylonger for a new AME value. Instead the ISOLTRAP value is now contributing with 8.3\% to the new atomic-mass evaluation, whereas the reaction $Q$-values of $^{26}$Mg(p,n)$^{26}$Al, $^{26}$Mg($^3$He,t)$^{26}$Al-$^{14}$N()$^{14}$O, and $^{42}$Ca($^3$He,t)$^{42}$Sc-$^{26}$Mg()$^{26}$Al are contributing with 78.4\%, 8.3\%, and 5.0\%, respectively (see fig.~\ref{Aluminium_26}\,(b)).
\\
Altogether five reaction $Q$-value measurements of the reactions $^{26}$Mg(p,$\gamma$)$^{27}$Al and $^{27}$Al(p,$\gamma$)$^{28}$Si have contributed to the mass value of $^{27}$Al. Together with some other reactions $Q$-values they are displayed in fig.~\ref{Aluminium_27}\,(a). The three independent data points of the reaction $^{26}$Mg(p,$\gamma$)$^{27}$Al contributed 16.0\% and the two measurements of the reaction $^{27}$Al(p,$\gamma$)$^{28}$Si contributed 84.0\% to the mass value. 
The mass measurement of ISOLTRAP is contributing 20.0\% to the new average value (fig.~\ref{Aluminium_27}\,(b)). Therefore the influence of the reaction measurements of $^{26}$Mg(p,$\gamma$)$^{27}$Al and $^{27}$Al(p,$\gamma$)$^{28}$Si is reduced to 12.9\% and 67.2\%, respectively. All contributing data points agree within their uncertainties. Note that no $Q$-value measurement of the reaction $^{27}$Al(p,$\alpha$)$^{24}$Mg contributes to the mass value, even though the most precise data point has an uncertainty of only 0.21 keV~\cite{Maas1978}. However, this measurement has been rejected since it deviates by more than four standard deviations from the accepted mean value and disagrees with earlier published values by the same group.
\section{The masses of $^{38}$Ca and $^{39}$Ca}
\begin{figure}[h!]
    \centering
        \includegraphics[width=8cm]{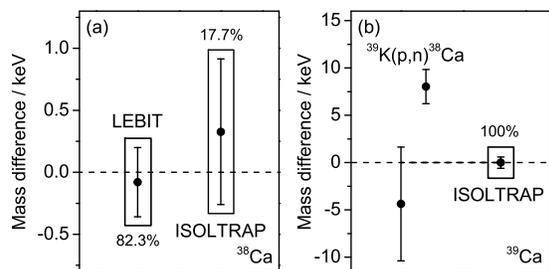}
        \caption{(a) Mass differences of $^{38}$Ca derived from two Penning trap measurements by the LEBIT facility~\cite{Boll2006} and ISOLTRAP~\cite{Geor2007a}. Analog denotes the zero line the new value of the atomic-mass evaluation. (b) Mass difference of $^{39}$Ca derived from two reaction measurements compared to the Penning trap measurement of ISOLTRAP.}
         \label{Calcium_38_39}
\end{figure}

Until Bollen \emph{et al.} performed a mass measurement with the Penning trap mass spectrometer LEBIT~\cite{Boll2006}, the mass of $^{38}$Ca was determined from the most precise of three $Q$-value measurements of the reaction $^{40}$Ca(p,t)$^{38}$Ca. The uncertainty of the $^{38}$Ca mass value was 5 keV. The LEBIT measurement reduced the uncertainty by more than an order of magntitude to 0.28\,keV and lowered the value by 3.3\,keV. 
We confirmed this new value by measuring the mass of $^{38}$Ca at ISOLTRAP. Our mass uncertainty is 0.59\,keV~\cite{Geor2007a}, whereas both values differ only by 0.43\,keV. The reason for our larger uncertainty despite using the favorable Ramsey excitation technique is the factor of five higher cyclotron frequency at MSU, since they measured doubly charged $^{38}\mbox{Ca}^{2+}$ ions in a 9.4\,T magnetic field while we measured the singly charged, heavier molecular sideband $^{38}\mbox{Ca}^{19}\mbox{F}^+$ in a 5.9\,T magnetic field. The higher frequency leads to a smaller relative uncertainty, since the absolute uncertainty of the frequency determination is identical. In the new atomic-mass evaluation the ISOLTRAP value contributes nevertheless 17.7\%. The main contribution of 82.3\% comes from the LEBIT measurement (see fig.~\ref{Calcium_38_39}\,(a)). In total the mass value is increased by 0.08\,keV compared to the LEBIT value.
\\
Two $Q$-value measurements of the reaction $^{39}$K(p,$\gamma$)$^{39}$Ca have so far been performed (see fig.~\ref{Calcium_38_39}\,(b)). Despite their relative small uncertainties of 1.8 keV~\cite{Rao1978} and 6 keV~\cite{Kemp1970}, they differ by more than 12 keV. Until now the value of Rao \emph{et al.}~\cite{Rao1978} was used as the AME value. The new ISOLTRAP value disagrees with this value by more than four standard deviations, but is in perfect agreement with the value of Kemper \emph{et al.}~\cite{Kemp1970}. Since their uncertainty is a factor of ten larger, the ISOLTRAP value determines in the new atomic-mass evaluation the value by 100\%. 

\section{Conclusion}
We have demonstrated that separated oscillatory fields are an excellent tool for high-precision mass measurements. The final mass values are limited by our systematic uncertainties, since statistical uncertainties of less than $2\cdot 10^{-9}$ have been achieved. This shows the potential for future mass measurements on the level of $10^{-9}$, which is especially important for tests of the CVC hypothesis, as discussed above. Therefore, it is required to reduce significantly the presently limiting systematic uncertainties. Because the measurement is up to ten times faster, the Ramsey technique is of special interest for short-lived radionuclides which are generally produced with very low yield. Here, further improvements by optimizing the frequency scan range and step size are presently in progress. For longer excitation cycles the uncertainty of the frequency determination is expected to be dominated by the statistical uncertainty of the measured time of flight, when the number of recorded ions is kept constant. A further detailed investigation is ongoing. Penning trap mass measurements on superallowed $\beta$-emitters are needed to test and improve older reaction-based $Q$-value measurements. Here, the masses of $^{26,27}$Al and $^{38}$Ca have been confirmed and for $^{39}$Ca a new mass value has been established.




\acknowledgments
This work was supported by the German BMBF under contracts 06GF151, 06GF186I and 06MZ215, by the EU under contracts HPMT-CT-2000-00197 
(Marie Curie Fellowship) and RII3-CT-2004-506065 (TRAPSPEC), by the European Union Sixth Framework through RII3-EURONS (contract no. 506065), and by the Helmholtz association of national research centers (HGF) under contract VH-NG-037. We also thank the ISOLDE 
Collaboration as well as the ISOLDE technical group for their assistance.

\end{document}